\documentclass[12pt]{article}
\usepackage{graphicx}
\usepackage{xspace}

\def\pbnr{}
\def\speaker{Ed Greening}
\def\onbehalfof{the \textup{LHCb} collaboration}
\def\title{Rare Charm Decays at LHCb}
\def\affiliation{The University of Oxford}
\def\support{The workshop was supported by the University of Manchester, IPPP, STFC, and IOP}

\newcommand\pubnumber{\pbnr}
\newcommand\pubdate{\today}
\def\Title#1{\begin{center} {\Large #1 } \end{center}}
\def\Author#1{\begin{center}{ \sc #1} \end{center}}

\newcommand{\OnBehalf}[1]{\sbox0{#1}\ifdim\wd0=0pt
        {}
	\else
	{\\on behalf of #1}
	\fi}
\newcommand{\SupportedBy}[1]{\sbox0{#1}\ifdim\wd0=0pt
        {}
	\else
	{\footnote{#1}}
	\fi}
\def\Address#1{\begin{center}{ \it #1} \end{center}}

\newcommand\pubblock{\includegraphics[width=5cm]{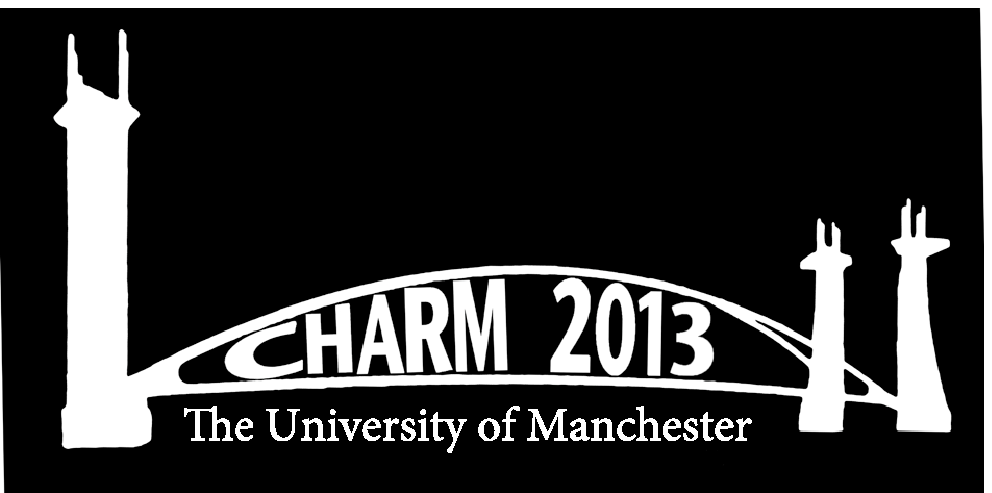}\hfill{\begin{tabular}{l} \pubnumber\\
         \pubdate  \end{tabular}}}
\newenvironment{Abstract}{\begin{quotation}  }{\end{quotation}}
\newenvironment{Presented}{\begin{quotation} \begin{center} 
             PRESENTED AT\end{center}\bigskip 
      \begin{center}\begin{large}}{\end{large}\end{center} \end{quotation}}

\def\venue{The 6$^{th}$ International Workshop on Charm Physics\\
(CHARM 2013)\\
Manchester, UK,  31 August -- 4 September, 2013}

\def\beq{\begin{equation}}
\def\eeq#1{\label{#1}\end{equation}}
\def\eeqn{\end{equation}}

\def\beqa{\begin{eqnarray}}
\def\eeqa#1{\label{#1}\end{eqnarray}}
\def\eeqan{\end{eqnarray}}

\let\bar=\overbar

\def\D{{\cal D}}

\def\Dslash{\not{\hbox{\kern-4pt $D$}}}
\def\dslash{\not{\hbox{\kern-2pt $\del$}}}

\def\BR{\mbox{\rm BR}}

\def\msb{{\bar{\ssstyle M \kern -1pt S}}}

\def\D  		 {\ensuremath{D}\xspace}
\def\Dp  		 {\ensuremath{D^{+}}\xspace}
\def\Ds  		 {\ensuremath{D_{s}^{+}}\xspace}
\def\Dps  		 {\ensuremath{D_{(s)}^{+}}\xspace}

\def\Dz  		  {\ensuremath{D^{0}}\xspace}
\def\Dstar            {\ensuremath{D^{+*}}\xspace}

\def\mup              {\ensuremath{\mu^{+}}\xspace}

\def\mumu           {\ensuremath{\mu^{+} \mu^{-}}\xspace}

\def\pip                {\ensuremath{\pi^{+}}\xspace}
\def\pim               {\ensuremath \pi^{-}\xspace}
\def\pipi               {\ensuremath \pi^{+}\pi^{-}\xspace}
\def\pis                {\ensuremath{\pi^{+}_{s}}}

\def\B               {\ensuremath{B}\xspace}

\def\K               {\ensuremath{K}\xspace}
\def\Kp               {\ensuremath{K^{+}}\xspace}
\def\Km               {\ensuremath{K^{-}}\xspace}
\def\KK               {\ensuremath{ \Kp \Km }\xspace}

\def\V               {\ensuremath{V}\xspace}

\def\Peta          {\ensuremath{ \eta }\xspace}
\def\Prho          {\ensuremath{ \rho^{0} }\xspace}
\def\Pomega     {\ensuremath{ \omega }\xspace}
\def\Prhomega  {\ensuremath{ \Prho / \Pomega }\xspace}
\def\Pphi          {\ensuremath{ \phi }\xspace}

\def\DstarDecay  {\ensuremath{\Dstar \to \Dz \pip}\xspace}

\def\twobody 	  {\ensuremath{\Dz \to \mumu}\xspace}

\def\twobodypipi {\ensuremath{\Dz \to \pipi}\xspace}
\def\twobodykpi  {\ensuremath{\Dz \to \Km \pip}\xspace}
\def\twobodySL  {\ensuremath{\Dz \to \pim \mup \nu_{\mu}}\xspace}

\def\threebody   	{\ensuremath{\Dps \to \pip\mumu}\xspace}

\def\threebodyD        {\ensuremath{\Dp \to \pip\mumu}\xspace}
\def\threebodyDs      {\ensuremath{\Ds \to \pip\mumu}\xspace}

\def\threebodyres        {\ensuremath{\Dps \to \pip (V \to \mumu)}\xspace}

\def\threebodypipipi   {\ensuremath{\Dps \to \pip\pipi}\xspace}

\def\fourbody             {\ensuremath{\Dz \to \pipi\mumu}\xspace}
\def\fourbodyres       {\ensuremath{\Dz \to \pipi (V \to \mumu)}\xspace}
\def\fourbodyphi       {\ensuremath{\Dz \to \pipi (\phi \to \mumu)}\xspace}
\def\fourbodyCLEO  {\ensuremath{\Dz \to \KK\pipi}\xspace}

\def\to            {\ensuremath \rightarrow \xspace}
\def\invfb            {\ensuremath{\mathrm{\, fb^{-1}}}\xspace}
\def\BF                {\ensuremath{\cal B}}
\def\BR          {\ensuremath{\cal B}\xspace}
\def\dM               {\ensuremath{\Delta m}\xspace}
\def\mevcc         {\ensuremath{{\mathrm{\,Me\kern -0.1em V\!/}c^2}}\xspace}
\def\gevcc          {\ensuremath{{\mathrm{\,Ge\kern -0.1em V\!/}c^2}}\xspace}
\def\tev          {\ensuremath{{\mathrm{\,Te\kern -0.1em V\!}}}\xspace}
\def\sqs   	{\ensuremath{\protect\sqrt{s}} \xspace}

\begin{document}
\begin{titlepage}
\pubblock

\vfill
\Title{\title}
\vfill
\Author{\speaker\SupportedBy{\support}\OnBehalf{\onbehalfof}}
\Address{\affiliation}
\vfill
\begin{Abstract}
Studies of rare decays are an indirect probe of New Physics (NP).
This document presents recent measurements of rare decays in the charm sector by the LHCb experiment.
The analyses are performed with proton-proton collision data at \sqs = 7 \tev recorded in 2011.
\end{Abstract}
\vfill
\begin{Presented}
\venue
\end{Presented}
\vfill
\end{titlepage}
\def\thefootnote{\fnsymbol{footnote}}
\setcounter{footnote}{0}
%

\section{Introduction}


Flavour-changing neutral current (FCNC) processes are rare within the Standard Model (SM) as they cannot occur at tree level. At loop level, they are suppressed by the both the Glashow-Iliopoulos-Maiani (GIM)~\cite{Glashow:1970gm} and the Cabibbo-Kobayashi-Maskawa (CKM)~\cite{Cabibbo:1963, Kobayashi:1973} mechanisms but are nevertheless well established in processes that involve \K and \B mesons.
In contrast to the \B meson system, where the near-unity value of $|V_{ub}|$ and very high mass of the top quark in the loop weaken the suppression, the cancellation is almost exact in \D meson decays leading to lower SM branching fractions (\BR).
This suppression provides a unique opportunity to probe the effects of NP on the coupling of up-type quarks in electroweak processes.
NP models may introduce additional diagrams that {\it a priori} need not be suppressed in the same manner as the SM contributions.
Enhancement in the \BR of such decays would therefore be a sign of NP.

The large number of \D mesons created at the LHC and LHCb's excellent ability to discriminate between pions and muons~\cite{Alves:2008, Adinolfi:2013} 
mean that the detector is in a outstanding position to investigate rare charm decays.

\section{\twobody}
The decay \twobody is very rare in the SM because of additional helicity suppression.
The short distance perturbative contribution to the \BR is of the order of $10^{-18}$ while the long distance non-perturbative contribution, 
dominated by the two-photon intermediate state, is estimated to be of the order $10^{5}$ higher~\cite{Burdman:2001tf}.

A search for the decay is performed with 0.9 \invfb of data~\cite{Aaij:2013cza}. By taking the \Dz from \DstarDecay decays, a two-dimensional fit is performed in m(\mumu) and \dM ($\equiv m(\pip\mumu) - m(\mumu)$).
The measured \BR is normalised with the decay \twobodypipi.
Peaking backgrounds from the misidentified hadronic decays \twobodypipi and \twobodykpi and the misidentified and partially reconstructed semileptonic decay \twobodySL are also taken into account.

The observed number of events is consistent with the background expectations and corresponds to an upper limit of \BF(\twobody) $< 6.2 \, (7.6) \times 10^{-9}$ at 90\% (95\%) CL.
This represents an improvement of more than a factor twenty with respect to previous measurements~\cite{Petric:2010} but remains several orders of magnitude larger than the SM prediction.


\section{\threebody}
The decay \threebody proceeds via short and long distance contributions. The long-distance contributions are mediated by intermediate resonances, \threebodyres, where \V $=$ \Pphi, \Peta, \Prho or \Pomega, whose large \BR mask any deviation from the much smaller non-resonant SM prediction, caused by NP.

A search for the decay is performed with 1.0 \invfb of data~\cite{Aaij:2013sua}. The data is binned in m(\mumu) allowing the long and short distance contributions to be separated. The binning definitions are shown in Table~\ref{tab:3bodyYields}.
The contribution from the intermediate \Prho and \Pomega resonances are grouped together as it is non-trivial to separate them.
The signal yields in each bin are determined with a simultaneous fit to the m($\pi^+$\mumu) distribution of the m(\mumu) bins and shown in Table~\ref{tab:3bodyYields}.
The parameters of the shapes defining the \Dps signals are determined simultaneously across all bins.
Candidates from the kinematically similar \threebodypipipi decay form an important peaking background.
Data-driven methods are used to parameterise their contributions.

The observed data, away from resonant structures, is compatible with the background-only hypothesis, and no enhancement is observed.
Upper limits in the low and high m(\mumu) bins are calculated by normalising with the \Pphi resonances.
The upper limits in the low and high m(\mumu) bins, assuming a phase space \mumu distribution, are extrapolated across the entirety of m(\mumu) by taking into account the relative efficiencies in each bin.
Upper limits on the non-resonant signal component of \BF(\threebodyD) $< 7.3 \, (8.3) \times 10^{-8}$ and \BF(\threebodyDs) $< 4.1 \, (4.8) \times 10^{-9}$ at 90\% (95\%) CL are set.
These represent an improvement of a factor 50 with respect to the previous limits~\cite{Abazov:2007aj, Link:2003qp}, but \BF(\threebodyD) is still an order of magnitude larger than the SM prediction.

\begin{table}[htp]
\footnotesize
\centering
\caption{Signal yields for the \threebody fits.}
\begin{tabular}{cccc}
Bin description & m(\mumu) range [$\mevcc$] & \D yield & \Ds yield\\
\hline
low-m(\mumu) & $\phantom{1}250-\phantom{1}525$ & $\phantom{00}-3\pm11$ & $\phantom{-000}1\pm\phantom{0}6$\\
\Peta & $\phantom{1}525-\phantom{1}565$ & $\phantom{-00}29\pm\phantom{0}7$ & $\phantom{-00}22\pm\phantom{0}5$\\
\Prhomega & $\phantom{1}565-\phantom{1}850$ & $\phantom{-00}96\pm15$ & $\phantom{-00}87\pm12$\\
\Pphi & $\phantom{1}850-1250$ & $\phantom{-}2745\pm67$ & $\phantom{-}3855\pm86$\\
high-m(\mumu) & $1250-2000$ & $\phantom{-00}16\pm16$ & $\phantom{0}-17\pm16$\\
\end{tabular}
\label{tab:3bodyYields}
\end{table}

\section{\fourbody}
The non-resonant component of the decay \fourbody has an expected SM \BR of the order $10^{-9}$~\cite{Cappiello:2012vg}.
The branching fraction for these decays is expected to be dominated by long-distance contributions involving resonances, such as \fourbodyres, where \V can be any of the light mesons \Peta, \Prho, \Pomega or \Pphi.
The corresponding branching fractions can reach O($10^{-6}$)~\cite{Cappiello:2012vg}.

\begin{figure}[htbp]
\begin{center}
\includegraphics[width=0.49\textwidth]{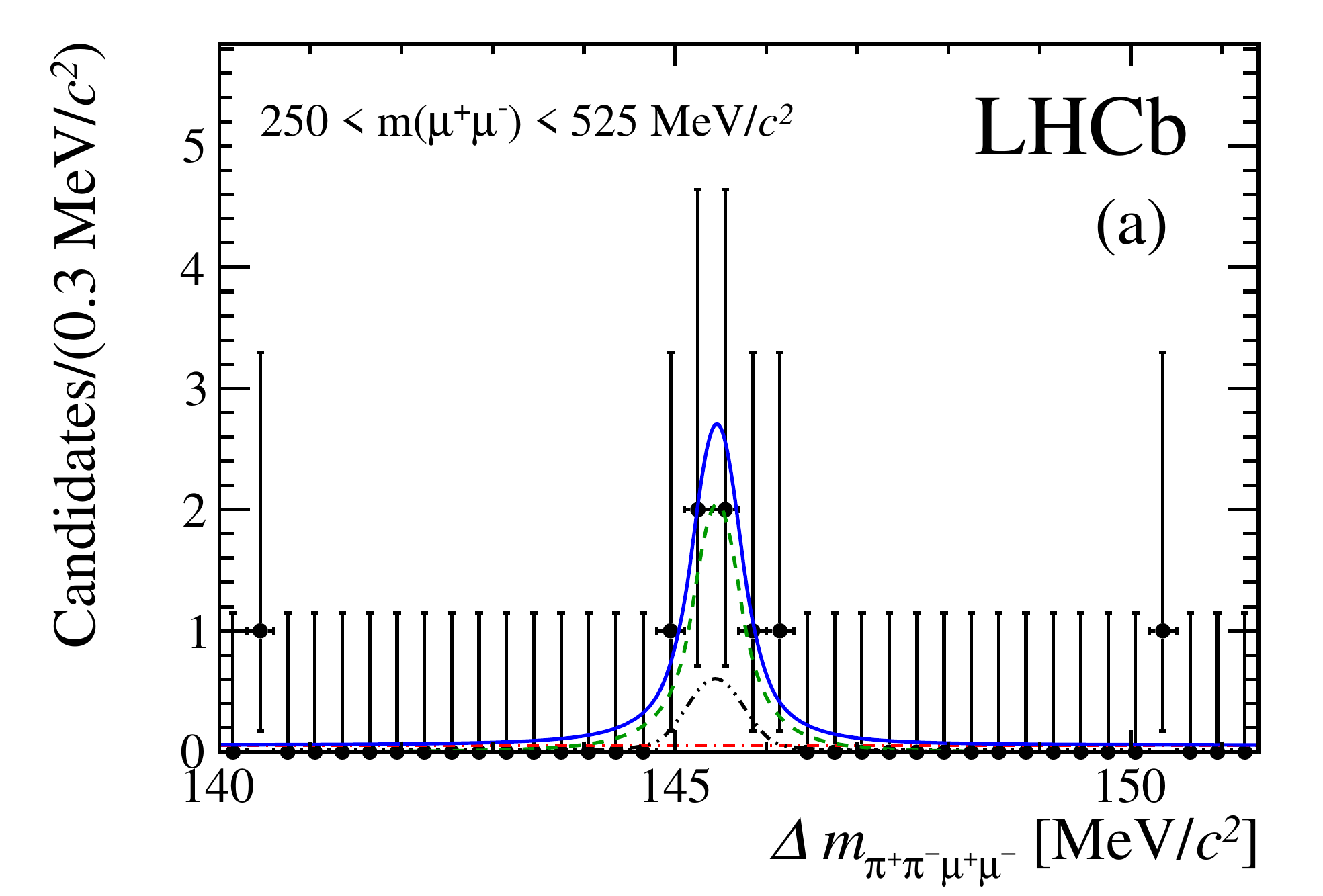}
\includegraphics[width=0.49\textwidth]{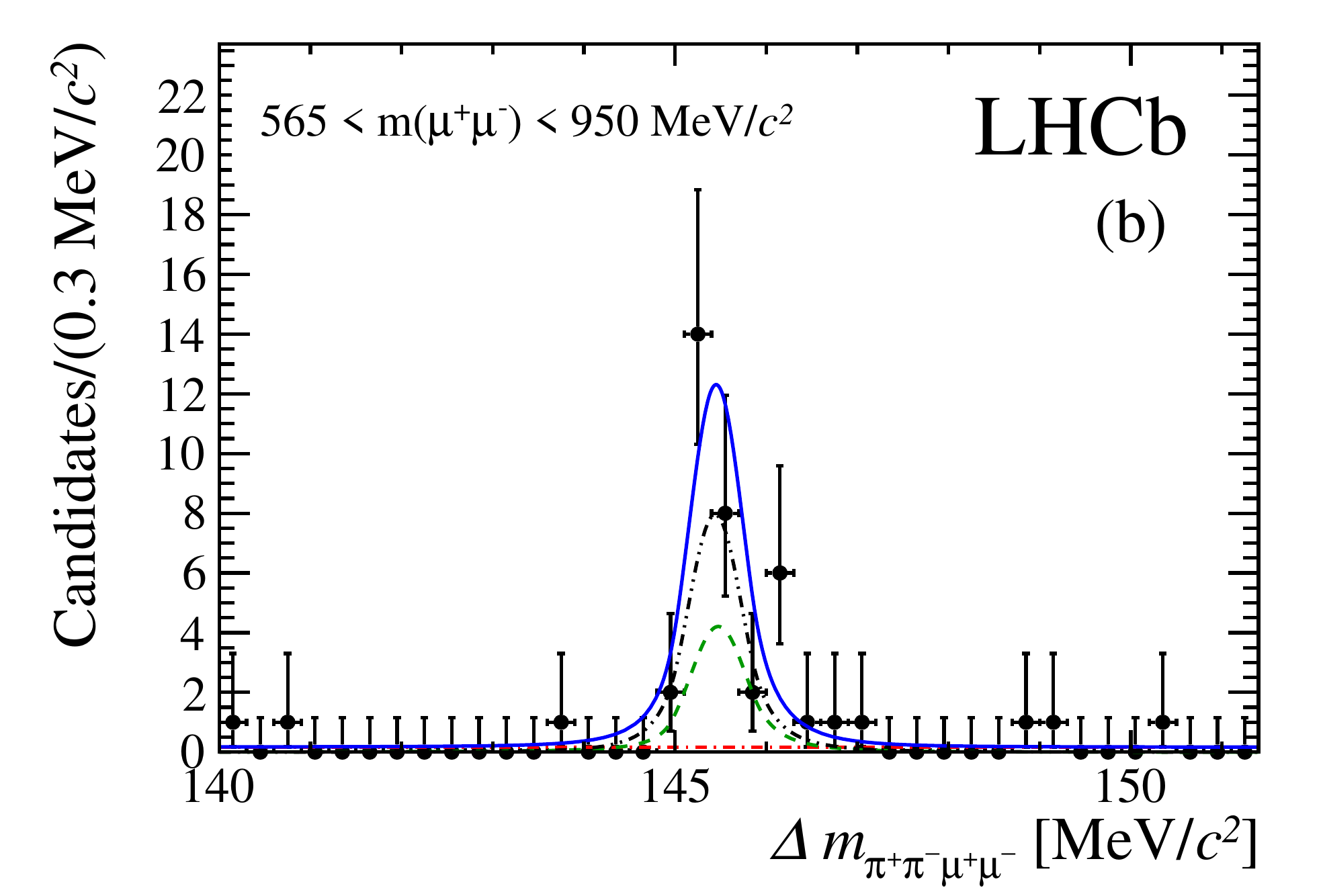}
\includegraphics[width=0.49\textwidth]{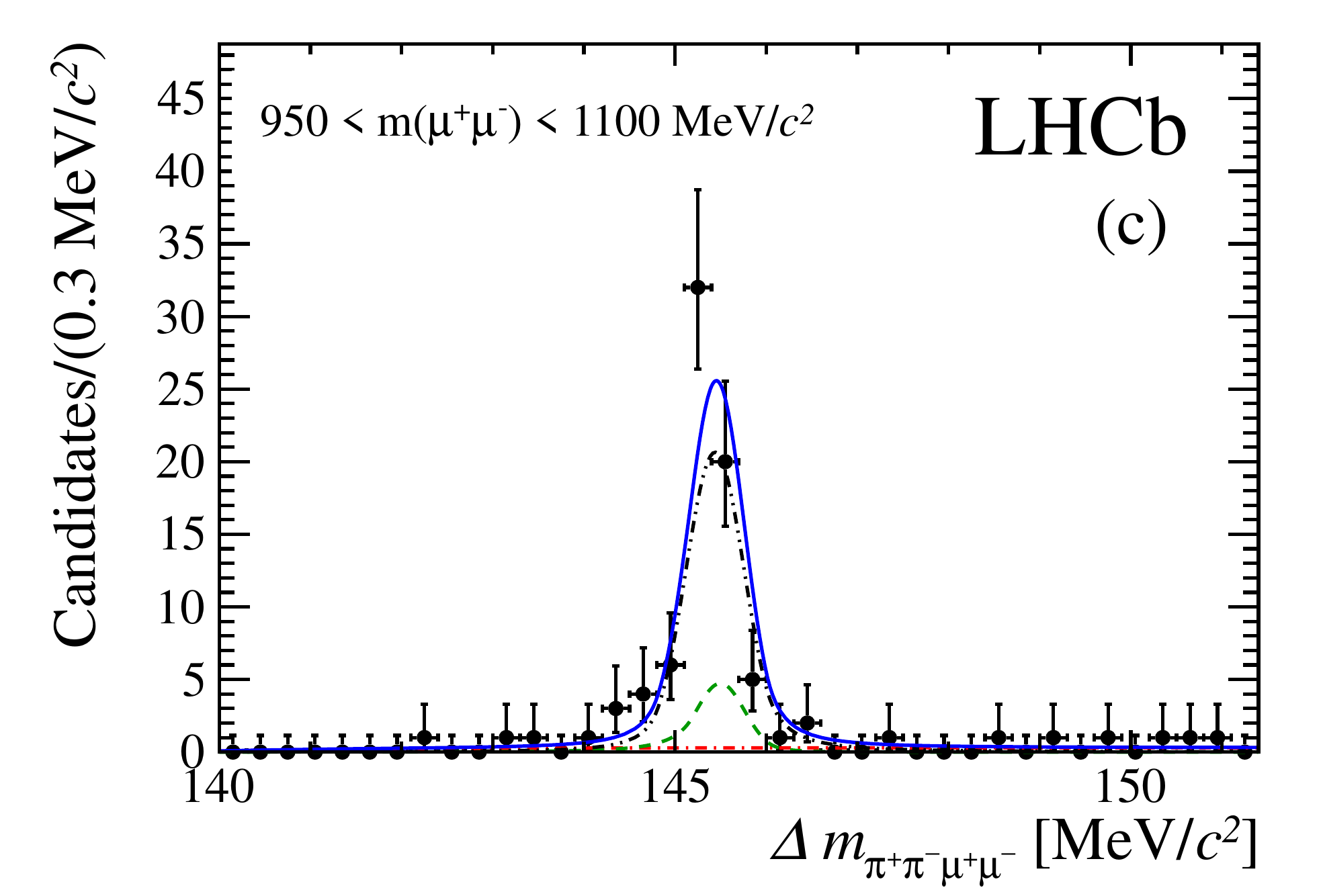}
\includegraphics[width=0.49\textwidth]{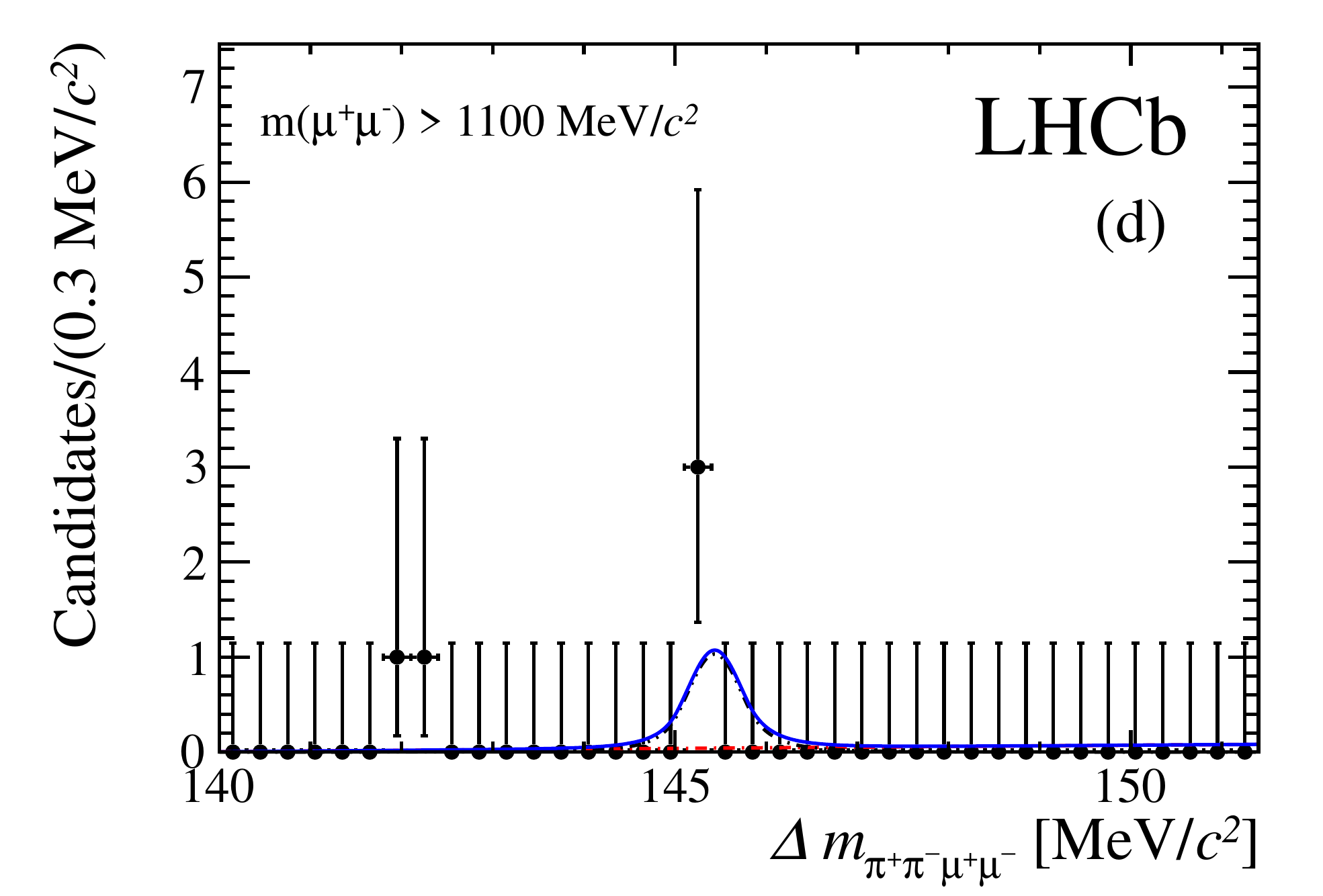}
\caption{Distributions of \dM for \fourbody decays in the (a) low-m(\mumu), (b) \Prhomega, (c) \Pphi, and (d) high-m(\mumu) bins, with the \Dz invariant mass in the range $1840-1888$ \mevcc. The data are shown as points (black) and the fit result (dark blue line) is overlaid. The components of the fit are also shown: the signal (black double-dashed double-dotted line), the peaking background (green dashed line) and the non-peaking background (red dashed-dotted
line).}
\label{fig:4bodyFitsDM}
\end{center}
\end{figure}

\begin{figure}[htbp]
\begin{center}
\includegraphics[width=0.49\textwidth]{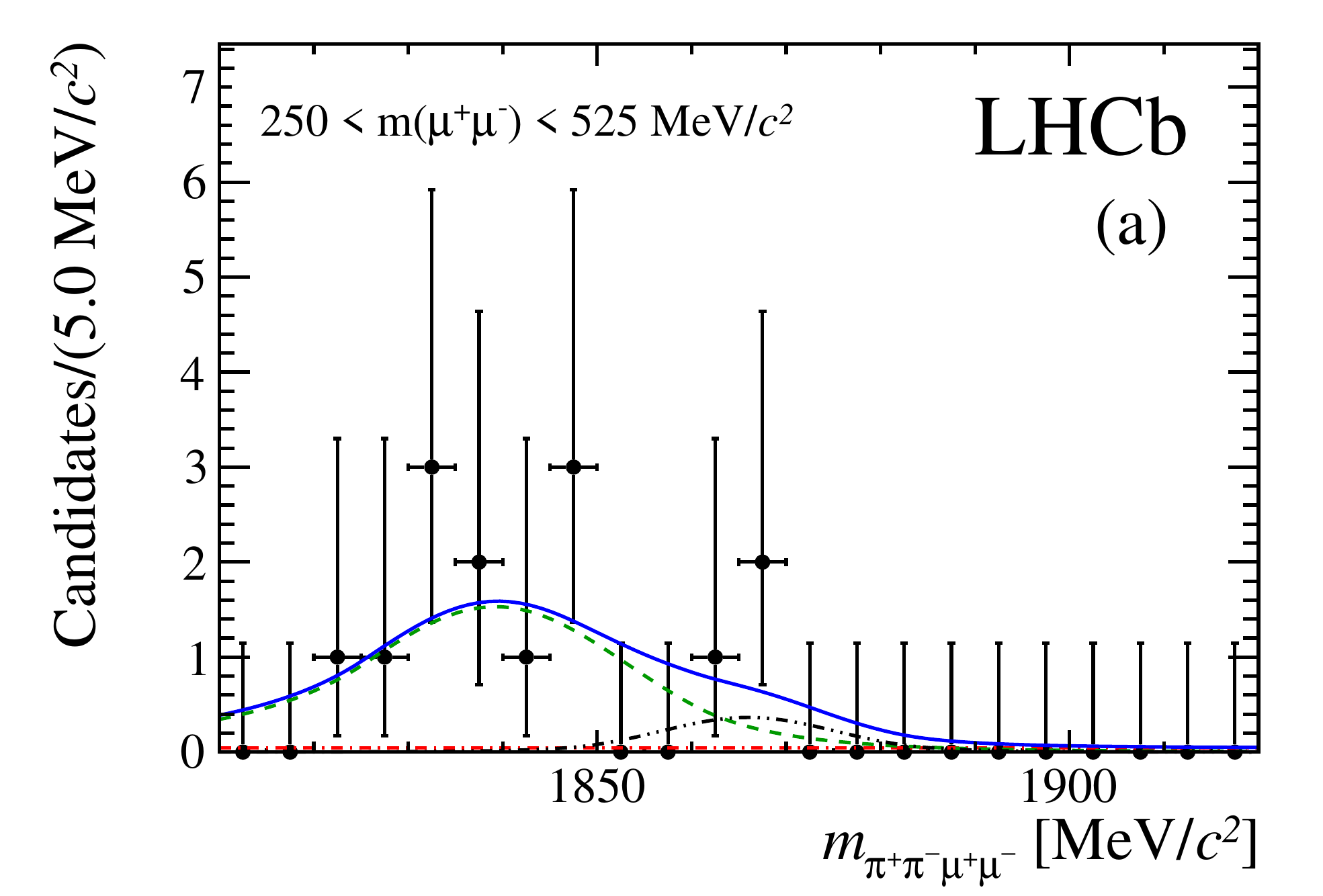}
\includegraphics[width=0.49\textwidth]{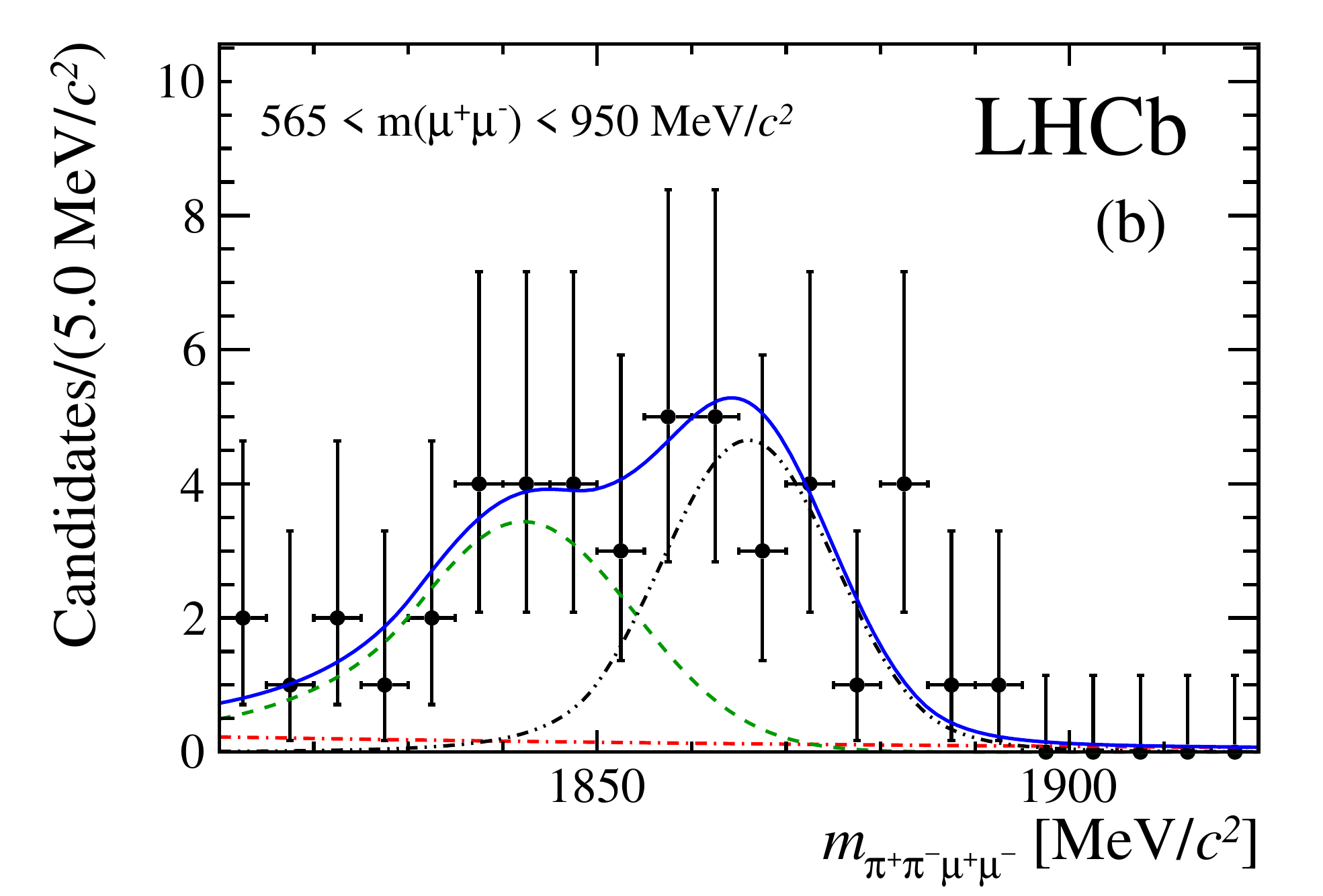}
\includegraphics[width=0.49\textwidth]{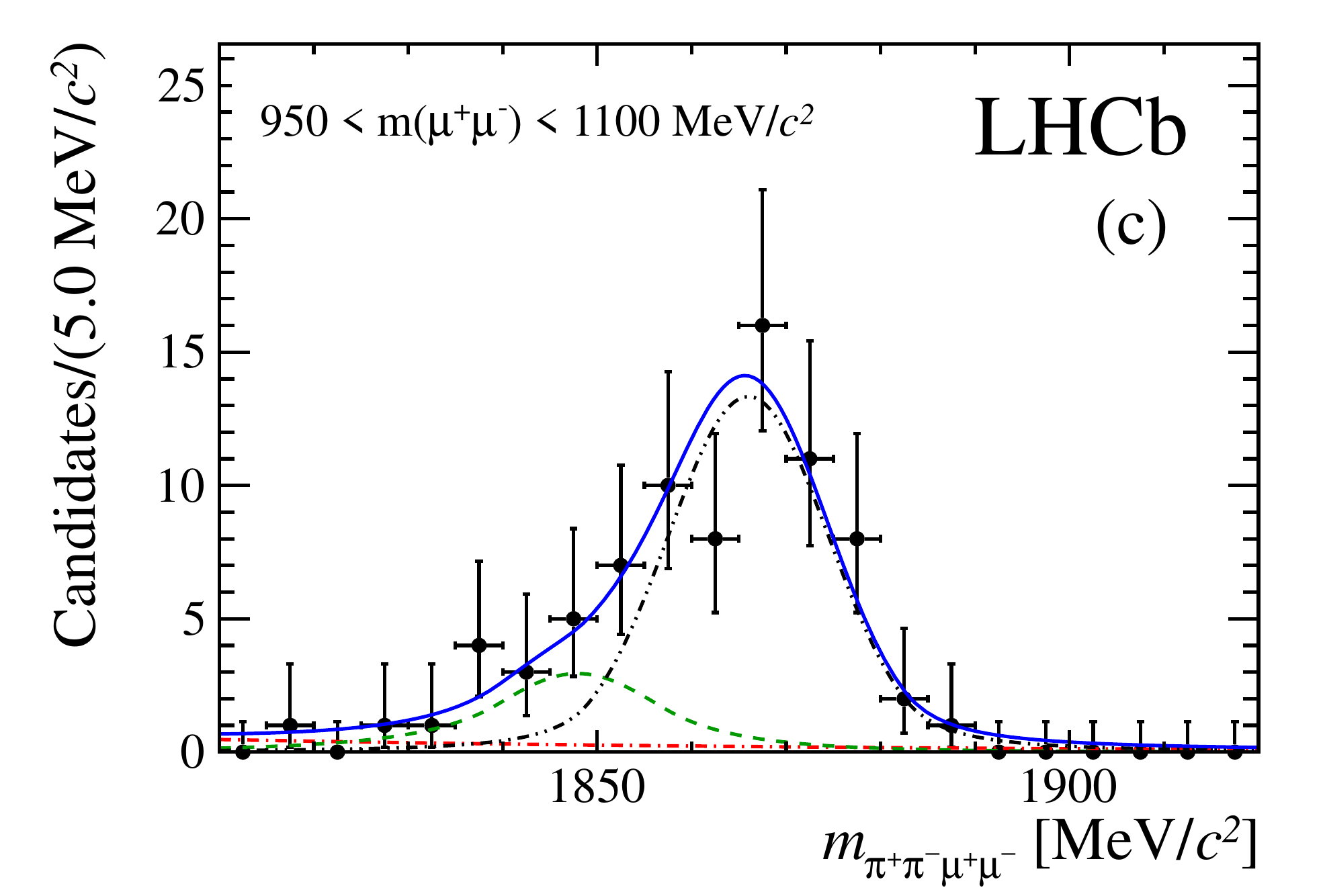}
\includegraphics[width=0.49\textwidth]{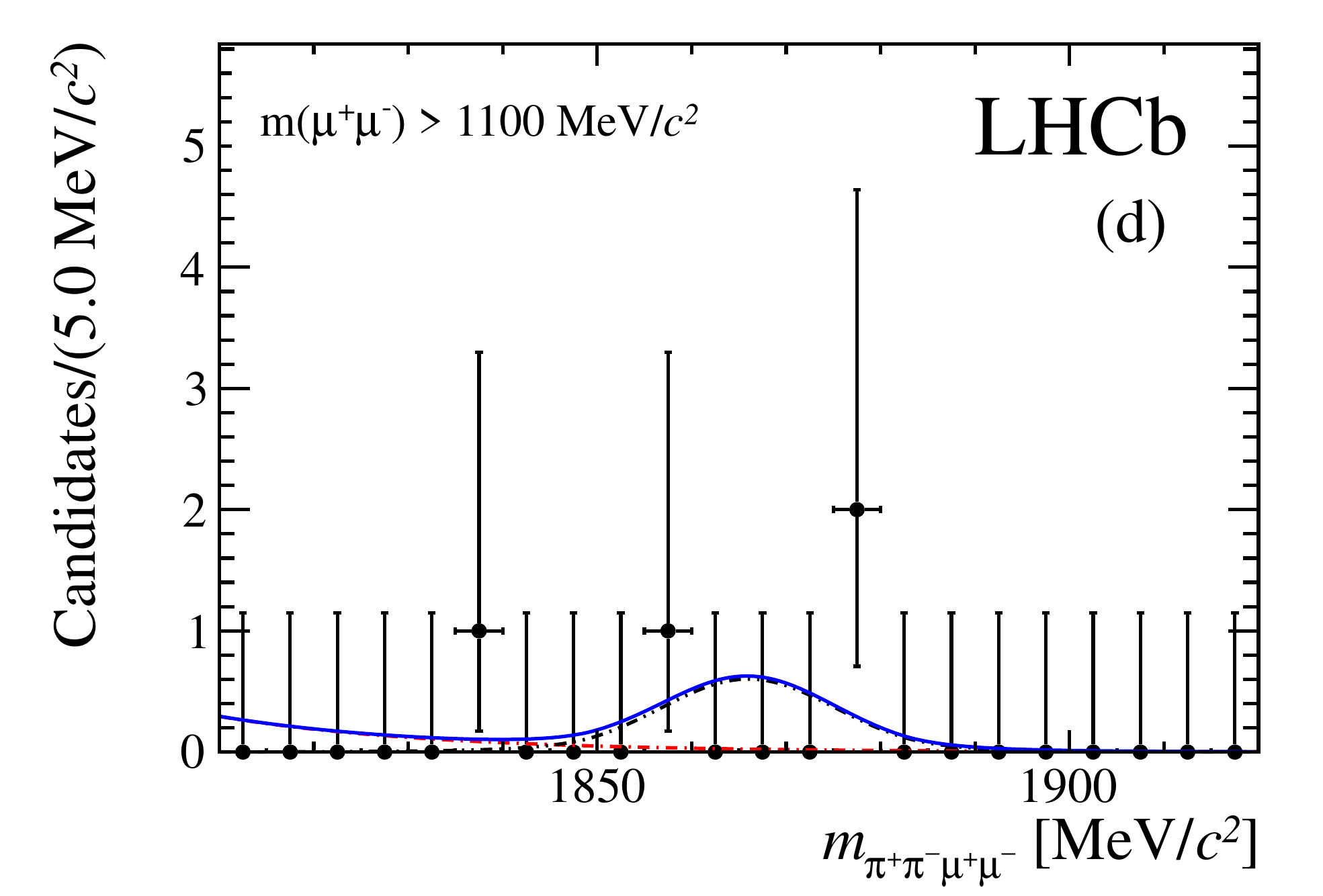}
\caption{Distributions of m($\pipi\mumu$) for \fourbody decays in the (a) low-m(\mumu), (b) \Prhomega, (c) \Pphi, and (d) high-m(\mumu) bins, with \dM in the range $144.4-146.6$ \mevcc. The data are shown as points (black) and the fit result (dark blue line) is overlaid. The components of the fit are also shown: the signal (black double-dashed double-dotted line), the peaking background (green dashed line) and the non-peaking background (red dashed-dotted
line).}
\label{fig:4bodyFitsM}
\end{center}
\end{figure}

A search for the decay is performed with 1.0 \invfb of data~\cite{Aaij:2013uoa}. 
The \fourbody data are split into four m(\mumu) bins: two bins containing the $\rho/\omega$ and $\phi$ resonances and two signal bins.
No \Peta bin is defined due to a lack of events after the analysis's offline selection.
The bin definitions are shown in Table~\ref{tab:4bodyYields}.
By taking the \Dz from \DstarDecay decays, a two-dimensional fit is performed in m(\mumu) and \dM ($\equiv m(\pis\pipi\mumu) - m(\pipi\mumu)$) in each bin.
$\Dz\to\pipi\pipi$ forms an important peaking background and data-driven methods are used to parameterise the contribution of this misidentified decay in each bin.
The \dM and m($\pipi\mumu$) fits can be seen in Fig.~\ref{fig:4bodyFitsDM} and Fig.~\ref{fig:4bodyFitsM}, respectively.
The fitted yields are provided in Table~\ref{tab:4bodyYields}.

The observed data, away from resonant structures, in both the low and high m(\mumu) bins, is compatible with the background-only hypothesis, and no enhancement is observed.
Upper limits in the low and high m(\mumu) bins are calculated by normalising with \BF(\fourbodyphi).
The normalisation \BR is estimated with the results of the amplitude analysis of the \fourbodyCLEO decay performed at CLEO~\cite{Artuso:2012df} and the known value of \BF(\Pphi\to\mumu)$/$\BF(\Pphi \to \KK)~\cite{PDG:2012}.
The upper limits in the low and high m(\mumu) bins, assuming a phase space \mumu distribution, are extrapolated across the entirety of m(\mumu) by taking into account the relative efficiencies in each bin.
An upper limit on the non-resonant signal component of \BF(\fourbody) $< 5.5 \, (6.7) \times 10^{-7}$ at 90\% (95\%) CL is set.

\begin{table}[htp]
\footnotesize
\centering
\caption{Signal yields for the \fourbody fits.}
\begin{tabular}{ccc}
Bin description & m(\mumu) range [$\mevcc$] & \fourbody yield\\
\hline
low-m(\mumu) & $\phantom{1}250-\phantom{1}525$ & $\phantom{0}2 \pm \phantom{0}2$\\
$\rho/\omega$ & $\phantom{1}565-\phantom{1}950$ & $23 \pm \phantom{0}6$\\
$\phi$ & $\phantom{1}950-1100$ & $63 \pm 10$\\
high-m(\mumu) & $>1100$ & $\phantom{0}3 \pm \phantom{0}2$\\
\end{tabular}
\label{tab:4bodyYields}
\end{table}

\section{Conclusion}
Before the second long shutdown of the LHC in 2017, LHCb expects to record an additional 5 \invfb at \sqs = 13 \tev.
This is in addition to the 1 and 2 \invfb of data at \sqs = 7 and 8 \tev, respectively, that LHCb already has on tape.
Together with anticipated improvements in LHCb's trigger system and analysis strategies, the higher centre-of-mass energy increases heavy flavour production cross-sections.
In comparison to the analyses detailed in this report, a factor of twenty increase in the number of observed decays can optimistically be hoped for.

A naive $\sqrt {20}$ scaling, would then give the following limits:
\BF(\twobody) $= 1 \times 10^{-8}$, an order of magnitude above the indirect bound;
\BF(\threebody) $= 2 \times 10^{-8}$, an order of magnitude above the SM expectation; and
\BF(\fourbody) $= 2 \times 10^{-7}$, two orders of magnitude above the SM expectation.
So although one would not expect to observe a SM signal before the LHCb upgrade, the phase space available to NP is set to be further probed.



\begin{thebibliography}{99}


\bibitem{Glashow:1970gm}
S.~L.~Glashow, J.~Iliopoulos and L.~Maiani,
Phys.\ Rev.\ D {\bf 2} (1970) 1285

\bibitem{Cabibbo:1963}
N.~Cabibbo,
Phys. Rev. Lett. 10, 531-533 (1963)

\bibitem{Kobayashi:1973}
M.~Kobayashi, T.~Maskawa,
Progress of Theoretical Physics 49 (2): 652-657 (1973)

\bibitem{Alves:2008}
A.~A.~Alves~Jr. {\it et al.}  [LHCb collaboration], 
JINST 3 (2008) S08005.

\bibitem{Adinolfi:2013}
M.~Adinolfi {\it et al.},
Eur. Phys. J. C73 (2013) 2431

\bibitem{Burdman:2001tf}
G.~Burdman, E.~Golowich, J.~L.~Hewett and S.~Pakvasa,
Phys.\ Rev.\ D {\bf 66} (2002) 014009

\bibitem{Aaij:2013cza}
  R.~Aaij {\it et al.}  [LHCb Collaboration],
  Phys.\ Lett.\ B {\bf 725} (2013) 15
  
\bibitem{Petric:2010}
  M.~Petric {\it et al.} [Belle collaboration],
  Phys. Rev. D81 (2010) 091102
  
\bibitem{Aaij:2013sua}
  R.~Aaij {\it et al.}  [LHCb Collaboration],
  Phys.\ Lett.\ B {\bf 724} (2013) 203
  
  \bibitem{Abazov:2007aj}
  V.~M.~Abazov {\it et al.}  [D0 Collaboration],
  Phys.\ Rev.\ Lett.\  {\bf 100} (2008) 101801
  
  \bibitem{Link:2003qp}
  J.~M.~Link {\it et al.}  [FOCUS Collaboration],
  Phys.\ Lett.\ B {\bf 572} (2003) 21
  
  \bibitem{Cappiello:2012vg}
  L.~Cappiello, O.~Cata and G.~D'Ambrosio,
  JHEP {\bf 1304} (2013) 135
  
  \bibitem{Aaij:2013uoa}
  R.~Aaij {\it et al.}  [LHCb Collaboration],
  LHCb-PAPER-2013-050 submitted to PLB
  
  \bibitem{Artuso:2012df}
  M.~Artuso {\it et al.}  [CLEO Collaboration],
  Phys.\ Rev.\ D {\bf 85} (2012) 122002
  
  \bibitem{PDG:2012}
  J. Beringer et al.  [Particle Data Group], 
  Phys. Rev. D86 (2012) 010001

\end{thebibliography}
\end{document}